\documentclass[11pt,a4paper]{article}

\usepackage{amsmath,amssymb,amsfonts}
\usepackage{hyperref}
\usepackage{geometry}
\usepackage{lipsum}
\usepackage{amssymb}
\usepackage{amssymb}
\usepackage{amsmath}
\usepackage{lipsum}

\usepackage[utf8]{inputenc}  
\usepackage{booktabs}        

\def\be{\begin{equation}}
\def\ee{\end{equation}}
\def\bea{\begin{eqnarray}}
\def\eea{\end{eqnarray}}
\title{\textbf{Detection of Axion Stars in Galactic Magnetic Fields}}

\author{
  Kuldeep J. Purohit$^{1}$ \and
  Jitesh R. Bhatt$^{2}$ \and
  Subhendra Mohanty$^{3}$ \and
  Prashant K. Mehta$^{4}$
}

\date{
\small
$^{1}$NSIT-IFSCS, Ahmedabad, India\\
$^{2}$Physical Research Laboratory, Ahmedabad, India\\
$^{3}$Indian Institute of Technology, Kanpur, India\\
$^{4}$The M.S. University of Baroda, Vadodara, India\\[1em]
\today
}

\begin{document}
\maketitle

\begin{abstract}
\noindent We perform a linear mode analysis of a uniformly distributed cloud of axion-like particles (ALPs) 
embedded in a magnetized intergalactic medium, in order to investigate the stability of axion stars 
under realistic astrophysical conditions. We find that when the frequency $\omega$ of transverse 
waves is much smaller than the collision frequency $\nu_c$ of the intergalactic plasma, the conversion 
of ALPs into photons occurs on timescales far longer than the age of the Universe, ensuring stability 
of the star. In the opposite regime, $\omega \gg \nu_c$, significant axion-to-photon conversion may 
occur if the condition $\tfrac{\beta^2}{m_a^2-\omega_p^2} < 1$ is satisfied, where $\beta$ depends 
on the ALP--photon coupling and the magnetic field, $m_a$ is the ALP mass, and $\omega_p$ is the plasma 
frequency. We have calculated up to  second order in perturbations to compute the effect of an ALP star. Since the calculated value of parameter $\beta ^2$
is extremely small in comparison with $\omega^2_p$, we argue that the direct detection of an axion star is highly unlikely in experiments like NCLE.

\noindent However, since the calculated $\beta$ is extremely small compared to $\omega_p$, this requires 
an unrealistically fine-tuned coincidence between $m_a$ and $\omega_p$.
As a consequence we argue that that detection of 
 Our results therefore suggest that axion stars remain stable in 
typical intergalactic environments, though extreme magnetic fields (e.g.\ near magnetars) may lead to different outcomes.
\end{abstract}







\section{Introduction}
\label{sec:intro}

QCD axions were originally proposed to solve the strong-$CP$ problem
\cite{Peccei, Steven, Wilczek}. They are also regarded as one of the most well-motivated candidates for cold dark matter (CDM) in cosmology. At the heart of the solution is the Peccei–Quinn 
mechanism \cite{Peccei}, which treats the $CP$-violating term in the QCD Lagrangian as a dynamical parameter that relaxes to zero and predicts the existence of a pseudoscalar boson called the axion. 
These axions can couple weakly with photons and other Standard Model particles through the Lagrangian density \cite{Peccei.38.1440, Raffelt.37.1237}:  

\begin{equation}
    \mathcal{L}= g_{a\gamma\gamma}\,{\bf E}\cdot{\bf B}\,a
\end{equation}

\noindent
where $g_{a\gamma\gamma}$ is the axion–photon coupling constant, 
${\bf E}$ and ${\bf B}$ denote the electric and magnetic field strengths, 
and $a$ is the axion field.  

Moreover, there are several other beyond-the-Standard-Model scenarios that require the existence of pseudoscalar particles called axion-like particles (ALPs). These particles also have similar forms of interaction as described by Equation (1), where $a$ now denotes the ALP field. The existence of ALPs is useful in providing potential explanations for various astrophysical phenomena such as the apparent transparency of the Universe to high-energy gamma rays \cite{PhysRevD.76.121301, PhysRevD.77.063001, PhysRevD.80.123511}, and the 3.55 keV galaxy cluster emission line \cite{ApJ789,PRL113(2014),PRD89}. ALPs are also considered to be well-motivated candidates for cold dark matter in the Universe [see review \cite{Choi_2021}].  

The electromagnetic interaction of axions or ALPs with Standard Model particles offers an interesting opportunity to observe them in experiments or astrophysical scenarios, e.g.\ “light shining through a wall” (LSW) experiments in a magnetic field [reviewed in \cite{arXiv:1011.3741}]. More advanced experimental searches are also expected \cite{Kotelnikov_2015, 1302.5647}. In addition, helioscope experiments such as CAST \cite{CAST2017} and the proposed IAXO \cite{Armengaud_2014} employ similar setups in which axion generation occurs in the Sun through photon–ALP and electron–ALP interactions.

There are also various cosmological scenarios that could lead to significant production of axions or ALPs in the early Universe. For example, it has been argued in \cite{Ahonen_1996} that axions produced during the QCD–hadron phase transition can survive for a very long time while undergoing electromagnetic interactions with other Standard Model particles. Similar results exist for ALPs as well \cite{Chavanis.023009, Gelmini_2023}. Given the expected abundance of dark matter, it is natural to expect that axions or ALPs can form gravitationally bound structures \cite{levkov, Purohit:2023izj}. Such stars can be regarded as a class of Bose stars.  

In the cosmology literature, many analytical and numerical studies describe the static and dynamical properties of Bose stars. It has been argued in \cite{levkov} that a Bose star made of ultralight dark matter with mass $m\sim 10^{-22}\,\mathrm{eV}$ may conceal a large fraction of dark matter from observation, potentially explaining the “missing satellite galaxies” problem in Milky Way simulations. Furthermore, if such a Bose star grows large, it may collapse or burst, emitting radio photons due to parametric resonance \cite{AstronLett12(1986)305}, which at suitable redshifts could explain fast radio bursts (FRBs) \cite{JETP}, or alternatively emit relativistic axions \cite{PRL118}.  

In the present work, we critically analyse the stability of a Bose star in the presence of intergalactic medium and magnetic fields. Since for ALPs the decay constant and mass can be treated as independent parameters, the discussion here is primarily applicable to ALP stars. For QCD axions, however, the mass $m_a$ and decay constant $f_a$ must satisfy the relation $m_a f_a \approx (10^8\,\mathrm{eV})^2$ [e.g.\ \cite{kyriazis2023diluteaxionstarsconverting}]. We make the following simplifying assumptions:  
(i) the Bose star made of ALPs is interspersed with the galactic magnetic field and intergalactic medium;  
(ii) within the star, ALPs, the magnetic field, and the intergalactic medium are taken to be homogeneous in space and time until perturbations are applied.  

This assumption facilitates Fourier-mode analysis to obtain the dispersion relation, although gravitational inhomogeneities would constrain the allowed wave-vector $k$ and star radius $R$ through the condition $kR \gg 1$. An estimate for the star radius can be obtained using the relation \cite{Visinelli:2017ooc, Chavanis.023009}
\begin{equation}
m_a R \sim \frac{M_{pl}}{f_a}, 
\end{equation}
where $M_{pl}=10^{18}\,\mathrm{GeV}$ is the reduced Planck mass and $f_a$ is the decay constant. Considering the axion window $10^9\,\mathrm{GeV} < f_a < 10^{11}\,\mathrm{GeV}$ \cite{Carena:2020ckf}, one obtains
\begin{equation}
R \sim \frac{2.4}{m_a}\,(10^9-10^{11})\,\mathrm{GeV}^{-1}.
\end{equation}

Before proceeding further, we note that Ref.\ \cite{kyriazis2023diluteaxionstarsconverting} studied a similar problem using the power formula
\begin{equation}
\frac{dP}{d\Omega} = \frac{\pi^4 (g_{a\gamma \gamma} \phi_0 \omega^2 R^2)^2}{32 k \omega}
     \left( \frac{\tanh\left(\frac{\pi k R}{2}\right)}{\cosh\left(\frac{\pi k R}{2}\right)} \right)^2 |B_0|^2,
\end{equation}
where $k$, $R$, and $B_0$ denote the magnitude of the wave-vector of the emitted radiation, the radius of the star, and the magnitude of the Milky Way magnetic field, respectively. Similar expressions have been used to study axion decay in neutron star environments [Refs, to be checked]. In the neutron star case, intergalactic plasma and hydrogen gas are absent.  

The author of \cite{kyriazis2023diluteaxionstarsconverting} argued that ALPs inside the star may decay into photons. The emitted photons subsequently generate transverse waves in the surrounding magnetized intergalactic plasma. 
 In particular, when the plasma frequency $\omega_p$ of the medium and the axion mass $m_a$ coincide, the star may be detectable with the NCLE instrument on board the Chang’e-4 satellite \cite{BENTUM2020856}. For $m_a \sim \omega_p \sim 10^{-12}\,\mathrm{eV}$, the star size is estimated to lie between $10^{-9}$ and $10^{-3}$ parsec . However, the use of the dispersion relation in this power formula is subject to restrictions, especially for $kR \sim 1$, where exponential power suppression can occur.  In addition, the dispersion relation used in the above power formula does not explicitly contain the possibility of converting ALP into photons.

In the present work, we instead perform a stability analysis based on the full dispersion relation, taking into account the intergalactic medium inside the star. In our view, this represents a more realistic scenario. The stability analysis presented here is based on various solutions of the exact dispersion relation.

\section{Methodology}
\label{sec:method}

One can write the total Lagrangian density for the axion--photon system as
\begin{equation}
\mathcal{L} = -\frac{1}{4}F^{\mu\nu}F_{\mu\nu} - J^\mu A_\mu + g_{a\gamma\gamma}\frac{1}{4}aF^{\mu\nu}\tilde{F}^{\mu\nu},
\label{eq:lagrangian}
\end{equation}
where $J^\mu$ is the electromagnetic current and $A^\mu$ is the gauge potential.  

The Euler--Lagrange equations yield a modified form of Maxwell’s equations in the presence of an axion field~\cite{SikiviePhysRevLett.51.1415, Ouellet}:
\begin{align}
\Box a + m_a^2 a &= g_{a\gamma\gamma}\, \mathbf{E} \cdot \mathbf{B}, \label{eq:axionEOM}\\
\nabla \cdot \mathbf{E} &= \rho - g_{a\gamma\gamma} \, \mathbf{B} \cdot \nabla a, \label{eq:maxwell1}\\
\nabla \cdot \mathbf{B} &= 0, \label{eq:maxwell2}\\
\nabla \times \mathbf{E} &= -\frac{\partial \mathbf{B}}{\partial t}, \label{eq:maxwell3}\\
\nabla \times \mathbf{B} &= \frac{\partial \mathbf{E}}{\partial t} + \mathbf{J} 
- g_{a\gamma\gamma}\left(\mathbf{E} \times \nabla a - \frac{\partial a}{\partial t} \mathbf{B}\right). \label{eq:maxwell4}
\end{align}

For a homogeneous system, one can neglect the spatial derivative terms in Eqs.~\eqref{eq:axionEOM}--\eqref{eq:maxwell4} to obtain

\begin{align}
\dot{\mathbf{E}} &= - g_{a\gamma\gamma} \, \mathbf{B} \, \dot{a} - \mathbf{J}, \label{eq:Ehom} \\
\ddot a + m_a^2 a &= g_{a\gamma\gamma} \, \mathbf{B} \cdot \mathbf{E}. \label{eq:ahom}
\end{align}

These equations are equivalent to those considered in Ref.~\cite{Ahonen_1996}, if one regards $\mathbf{J}=\sigma \mathbf{E}$ in the MHD approximation, with electrical conductivity
\begin{equation}
\sigma = \frac{\omega_p^2}{\nu_c},
\label{eq:conductivity}
\end{equation}
where $\omega_p$ is the plasma frequency and $\nu_c$ the collision frequency.

For linear analysis, any physical quantity is expanded as $A \sim A_0 + \delta A$, where $A_0$ is the background value while $\delta A$ denotes perturbations. Since the background electric field is absent and the magnetic field $\mathbf{B}_0$ is constant, Eqs.~\eqref{eq:maxwell1}--\eqref{eq:maxwell4} become
\begin{align}
\nabla \cdot \delta \mathbf{E} &= \delta \rho - g_{a\gamma\gamma} \, \mathbf{B}_0 \cdot \nabla \delta a - g_{a\gamma\gamma} \, \delta \mathbf{B} \cdot \nabla a_0, \label{eq:pert1}\\
\nabla \cdot \delta \mathbf{B} &= 0, \label{eq:pert2}\\
\nabla \times \delta \mathbf{E} &= -\frac{\partial \delta \mathbf{B}}{\partial t}, \label{eq:pert3}\\
\nabla \times \delta \mathbf{B} &= \frac{\partial \delta \mathbf{E}}{\partial t} + \delta \mathbf{J}
- g_{a\gamma\gamma}\left(\delta \mathbf{E} \times \nabla a_0 - \frac{\partial \, \delta a}{\partial t} \mathbf{B}_0\right). \label{eq:pert4}
\end{align}

With the plane-wave ansatz $\delta A \sim e^{-i\omega t + i\mathbf{k}\cdot\mathbf{x}}$, one rewrites the above equations as
\begin{align}
i\mathbf{k} \cdot \delta \mathbf{B} &= 0, \label{eq:pw1}\\
i\mathbf{k} \times \delta \mathbf{E} &= i\omega \, \delta \mathbf{B}, \label{eq:pw2}\\
i\mathbf{k} \times \delta \mathbf{B} &= -i\omega \, \delta \mathbf{E} + \delta \mathbf{J} - g_{a\gamma\gamma}\left(\delta \mathbf{E} \times \mathbf{k} a_0 + i\omega \, \delta a \, \mathbf{B}_0\right). \label{eq:pw3}
\end{align}

Considering a transverse electromagnetic wave with $\mathbf{k} \cdot \delta \mathbf{E} = 0$ and $\mathbf{k} \cdot \delta \mathbf{B} = 0$, Eq.~\eqref{eq:pw2} gives
\begin{equation}
 k_y \, \delta E_z =  \omega \, \delta B_x. 
 \label{eq:wave1}
\end{equation}
From Eq.~\eqref{eq:pw3}, one obtains
\begin{equation}
-i k_y \, \delta B_x \hat{z} = -i \omega \, \delta E_z \hat{z} + \delta J_z \hat{z} - g_{a\gamma\gamma}\left(-i \, \delta E_z k_y a \, \hat{x} + i \omega \, \delta a \, B_0 \, \hat{z}\right).
\label{eq:wave2}
\end{equation}

The $z$-component of Eq.~\eqref{eq:wave2} yields
\begin{equation}
-i k_y \, \delta B_x = -i \omega \, \delta E_z + \delta J_z - i \omega \, \delta a \, \beta,
\label{eq:wave3}
\end{equation}
where $\beta \equiv g_{a\gamma\gamma} B_0$.  

The axion field’s equation of motion becomes
\begin{equation}
(- \omega^2 + k_y^2 + m^2) \, \delta a = - g_{a\gamma\gamma} \, \delta E_z \, B_0.
\label{eq:axionPert}
\end{equation}

For the electron fluid, neglecting ions, the current is $J = -4\pi e n_e \, \mathbf{v}_e$. Linearizing, we obtain
\begin{equation}
m_e n_e (\partial_t \, \delta \mathbf{v}_e) = -e n_e (\delta \mathbf{E} + \delta \mathbf{v}_e \times \mathbf{B}_0) - m_e n_e \nu_c \, \delta \mathbf{v}_e.
\label{eq:fluid}
\end{equation}
This leads to the current response
\begin{equation}
J = i \frac{\omega_p^2}{\omega + i\nu_c} \, \mathbf{E},
\label{eq:current}
\end{equation}
with plasma frequency $\omega_p^2 = 4\pi e^2 n_e/m_e$. Hence,
\begin{equation}
\delta J_z = i \frac{\omega_p^2}{\omega + i\nu_c} \, \delta E_z.
\label{eq:Jz}
\end{equation}

Substituting Eqs.~\eqref{eq:wave1}, \eqref{eq:axionPert}, and \eqref{eq:Jz} into Eq.~\eqref{eq:wave3}, we arrive at the dispersion relation
\begin{equation}
\big[ (-\omega^2 + k^2)(\omega^2 - k^2 - m^2) + \beta^2 \omega^2 \big](\omega + i \nu_c) = -\omega_p^2 \, \omega (\omega^2 - k^2 - m^2).
\label{eq:disp1}
\end{equation}

Equivalently,
\begin{equation}
\left[ (-\omega^2 + k^2) + \frac{\omega_p^2 \, \omega}{\omega + i \nu_c} \right](\omega^2 - k^2 - m^2) = -\beta^2 \, \omega^2.
\label{eq:disp2}
\end{equation}

When $\beta=0$, Eq.~\eqref{eq:disp2} separates into independent branches: transverse plasma waves and free axion modes. A nonzero $\beta$ describes axion--photon conversion.  

\begin{table}[htbp]
    \centering
    \begin{tabular}{@{}ll@{}}
    \toprule
    \textbf{Parameter} & \textbf{Values (eV)} \\
    \midrule
    Plasma Frequency & $10^{-12}$ \\
    Conductivity & $10^{42}$ \\
    Collision Frequency & $10^{-25}$ \\
    $\beta$ & $10^{-26}$ \\
    Gyrofrequency & $10^{-15}$ \\
    \bottomrule
    \end{tabular}
    \caption{Interstellar Medium Parameters}
    \label{tab:interstellar_parameters}
\end{table}

\section{Results and Discussion}
\label{sec:results}

We first analyze two limiting regimes of the general dispersion relation in Eq.~\eqref{eq:disp2}: (i) the magnetohydrodynamic (MHD) limit $|\omega|\ll\nu_c$ and (ii) the non-MHD (collisionless) limit $|\omega|\gg\nu_c$~\cite{Boyd_Sanderson_2003}.

\paragraph{MHD limit $|\omega|\ll\nu_c$.}
Using Eq.~\eqref{eq:conductivity}, $\sigma=\omega_p^2/\nu_c$, Eq.~\eqref{eq:disp2} reduces to
\begin{equation}
\left[(-\omega^2+k^2) - i\,\sigma\,\omega\right](\omega^2-k^2-m^2) = -\beta^2\omega^2.
\label{eq:disp_MHD}
\end{equation}
For spatially homogeneous fluctuations ($k\to0$) one obtains
\begin{equation}
(\omega+i\sigma)(\omega^2-m^2)=\beta^2\omega,
\label{eq:disp_k0}
\end{equation}
which agrees with the MHD result in Ref.~\cite{Ahonen_1996}.

Introducing the dimensionless variables
\begin{equation}
 z\equiv \frac{\omega}{m},\qquad \epsilon\equiv \frac{\sigma}{m},\qquad \delta\equiv \frac{\beta^2}{m^2},
 \label{eq:def_zepsdelta}
\end{equation}
Eq.~\eqref{eq:disp_k0} becomes
\begin{equation}
 (z^2-1)(z+i\epsilon)=\delta\,z.
 \label{eq:cubic_z}
\end{equation}
Writing $z=x+i y$ and separating real and imaginary parts gives
\begin{align}
 x^2 &= \delta+3y^2+2\epsilon y+1, \label{eq:x2_expr}\\
 x^2(3y+\epsilon) &= y^3+y^2\epsilon+y+\epsilon+\delta y. \label{eq:imag_eq}
\end{align}
Eliminating $x^2$ between Eqs.~\eqref{eq:x2_expr} and \eqref{eq:imag_eq} yields a cubic equation for the damping rate $y$,
\begin{equation}
 8y^3+8\epsilon y^2+2y(\delta+\epsilon^2+1)+\delta\,\epsilon=0.
 \label{eq:cubic_y}
\end{equation}
A closed-form (Cardano) solution for the unique real root can be written, but in practice asymptotic limits are more transparent. For $\epsilon\gg\delta$ one finds the leading-order result
\begin{equation}
 y\simeq -\,\frac{\delta}{2\epsilon} = -\,\frac{\beta^2}{2\sigma\,m},
 \label{eq:y_asymp}
\end{equation}
consistent with Ref.~\cite{Ahonen_1996}. The damping rate is $\omega_I\equiv m\,y$ and the mode lifetime is $\tau\equiv |\omega_I|^{-1}$.

\paragraph{Non-MHD limit $|\omega|\gg\nu_c$.}
When collisions are negligible, Eq.~\eqref{eq:disp2} reduces to
\begin{equation}
 \left[(-\omega^2+k^2)+\omega_p^2\right](\omega^2-k^2-m^2)=-\beta^2\omega^2.
 \label{eq:disp_nonMHD}
\end{equation}

\paragraph{Conductivity and scale estimates.}
For estimates we use Gaussian units consistently with Sec.~\ref{sec:method}:
\begin{equation}
 \sigma=\frac{\omega_p^2}{\nu_c},\qquad \omega_p^2=\frac{4\pi e^2 n_e}{m_e},\qquad \nu_c\simeq n_e\,\sigma_c\,v_{\rm th},\quad v_{\rm th}=\sqrt{\frac{2T}{m_e}}.
 \label{eq:sigma_defs}
\end{equation}
Approximating the Coulomb (Rutherford) cross section with thermal averaging,
\begin{equation}
 \sigma_c\simeq \frac{\pi\alpha^2}{2\,\mathrm{KE}^2},
 \label{eq:rutherford}
\end{equation}
leads to the familiar $T^{3/2}$ scaling (Spitzer-like conductivity),
\begin{equation}
 \sigma\;\propto\; \frac{T^{3/2}}{\sqrt{m_e}}. \label{eq:spitzer_scaling}
\end{equation}
For a cold intergalactic medium (IGM) with $T\sim10\,\mathrm{K}$ and $n_e\sim 3\times10^{-2}\,\mathrm{cm}^{-3}$ one obtains a very small conductivity in energy units, $\sigma\sim10^{-5}\,\mathrm{eV}$ (order-of-magnitude estimate).

\section{Perturbative Expansion up to Second Order}
 In the previous section, we have calculated the dispersion relation up to a linear order in perturbation considering $\nu_c$ and $\beta^2$ as a small parameter. Calculated values of $\nu^2_c\approx \beta^2$ for the ALP star in the intergalactic medium, in this section, we provide upto a second-order correction to the dispersion relation for a more accurate solution of the dispersion relation.


The expansion reads
\begin{equation}
\omega = \omega_0 + \omega_1 + \omega_2 + \cdots ,
\end{equation}
with order counting
\begin{equation}
\omega_1 = \mathcal O(\nu_c),
\qquad
x_2 = \mathcal O(\beta^2,\nu_c^2).
\end{equation} 

\subsection{Zeroth-Order Solutions}

At zeroth order, setting $\nu_c = \beta = 0$, the dispersion relation yields the roots
\begin{equation}
    \omega_0 \in \left\{0,\; \pm \sqrt{k^2+\omega_p^2},\; \pm \sqrt{k^2+m^2}\right\},
\end{equation}
corresponding respectively to the static solution, the photon-like plasma branch, and the axion-like branch.

\subsection{First-Order Corrections}

Expanding to first order in $\gamma$, one obtains

\begin{equation}
    \omega_1 \in \left\{
    -\frac{i\nu_ck^2}{k^2+\omega_p^2},\;
    -\frac{i\nu_c \omega_p^2}{2(k^2+\omega_p^2)},\;
    -\frac{i\nu_c \omega_p^2}{2(k^2+\omega_p^2)},\;
    0,\;
    0
    \right\}.
    \label{eq:first_order}
\end{equation}
Thus, damping proportional to $\gamma$ appears for the photon-like modes, while the axion-like modes remain undamped at first order.

\subsection{Second-Order Corrections}

At second order, the correction $\omega_2$ reads
\begin{multline}
    \omega_2 \in \Bigl\{
    0,\;
    -\dfrac{\beta^2 \sqrt{k^2+\omega_p^2}}{2 (\omega_p^2-m^2)}
    + \dfrac{\nu_c^2 \omega_p^2(4k^2+\omega_p^2)}{8(k^2+\omega_p^2)^{5/2}},\; \\
    \dfrac{\beta^2 \sqrt{k^2+\omega_p^2}}{ (\omega_p^2-m^2)}
    - \dfrac{\nu_c^2 \omega_p^2(4k^2+\omega_p^2)}{8(k^2+\omega_p^2)^{5/2}},\;
    \dfrac{\beta^2\sqrt{k^2+m^2}}{2(\omega_p^2-m^2)},\;
   -\dfrac{\beta^2\sqrt{k^2+m^2}}{2(\omega_p^2-m^2)}
    \Bigr\}.
\end{multline}
    \label{eq:second_order}
\subsection{Discussion}

The perturbative expansion elucidates the interplay between plasma effects, collisional damping, and axion--photon mixing in a systematic manner. The structure of the corrections reveals a clear hierarchy of physical processes and highlights the regimes where non-perturbative phenomena are expected to emerge.  

At zeroth order ($\nu_c=\beta=0$), the dispersion relation factorizes into independent sectors: a trivial static solution, a photon-like plasma mode with frequency $\sqrt{k^2+\omega_p^2}$, and an axion-like mode with frequency $\sqrt{k^2+m^2}$. This spectrum provides the natural reference point for analyzing perturbative corrections.  

At first order in the small parameters, only the collisional term contributes. As shown in Eq.~\eqref{eq:first_order}, the photon-like modes acquire purely imaginary shifts proportional to $\nu_c$, describing collisional damping of electromagnetic excitations in the medium. By contrast, the axion-like modes remain undamped at this order, reflecting the absence of direct axion--plasma interactions. Thus, to leading order, damping is confined to the electromagnetic sector.  

At second order, both mixing ($\beta^2$) and higher-order collisional ($\nu_c^2$) effects become relevant [Eq.(4.5)]. The $\beta^2$ terms shift the real parts of the eigenfrequencies, producing the characteristic level repulsion between photon- and axion-like branches. This behavior is analogous to the avoided crossings in coupled oscillator systems and directly encodes axion--photon mixing. The sign structure indicates that one branch is shifted upward, while the other is shifted downward, the magnitude of the splitting set by the coupling strength $\beta$.  

The denominators $(\omega_p^2 - m^2)$ highlight the special role of resonance. When the plasma frequency approaches the axion mass, the perturbative expansion ceases to be reliable: the frequency shifts formally diverge, reflecting the onset of strong mixing. In this regime, axion--photon conversion is resonantly enhanced and must be treated with non-perturbative methods. However, the resonance condition also introduces a fine-tuning issue. The plasma frequency is determined by the local electron density, and achieving $\omega_p \simeq m$ requires highly specific values of $n_e$. Unless the medium naturally spans a wide range of densities, the probability of realizing such a condition is small, and efficient conversion becomes contingent on finely tuned environments. In realistic plasmas, spatial gradients in density can broaden the resonance and mitigate this sensitivity, enabling adiabatic conversion akin to the MSW effect for neutrinos. Nevertheless, the perturbative results make explicit that away from resonance mixing is parametrically suppressed, while near resonance the system is controlled by dynamics beyond perturbation theory.

\paragraph{Parameter hierarchies and implications.}
In the galactic/IGM environments considered here one typically has $\beta\ll\nu_c\ll\omega_p$ (using, e.g., $g_{a\gamma\gamma}\sim10^{-12}\,\mathrm{GeV}^{-1}$ and $B_0\sim\mu\mathrm{G}$ $\Rightarrow$ $\beta\sim10^{-29}\,\mathrm{eV}$). Therefore, the expansion leading to Eqs.~\eqref{eq:y_asymp}, \eqref{eq:first_order} and \eqref{eq:second_order} is well justified for most of parameter space. Situations with extremely strong fields (e.g. magnetars) would require revisiting the hierarchy and re-evaluating $\beta$ and $\sigma$ accordingly before applying Eq.~\eqref{eq:cubic_y}.

\section{Summary and Conclusions}
\label{sec:conclusion}

In this work, we have analyzed the stability of axion stars composed of axion-like particles (ALPs) 
in the presence of an intergalactic medium and magnetic fields. The question of stability was 
addressed using a linear Fourier mode analysis. This method imposes the constraint that the 
gravitational inhomogeneity parameter $R$ and the wave-vector $k$ satisfy $kR \gg 1$.  

The general dispersion relation derived in this work involves the plasma frequency $\omega_p$, 
the electron--ion collision frequency $\nu_c$, and the axion--photon coupling with the magnetic 
field $\beta$. For intergalactic environments, typical values of $\nu_c$ and $\beta$ are much 
smaller than $\omega_p$, allowing the use of perturbative techniques to obtain solutions.  

In the magnetohydrodynamic limit ($\omega \ll \nu_c$), we demonstrated that the axion decay rate 
is so small that the corresponding timescale is much longer than the age of the Universe. This 
ensures that axion stars remain stable in this regime. In the opposite limit ($\omega \gg \nu_c$), 
we found that significant ALP-to-photon conversion is possible in principle if the condition
\begin{equation}
\frac{\beta^2}{m_a^2-\omega_p^2} \sim 1
\end{equation}
is satisfied. However, since $\beta^2 \sim 10^{-56}\,\text{eV}^2$ and $\omega_p \sim 10^{-12}\,\text{eV}$ 
for intergalactic plasma, this condition requires extremely fine tuning between $m_a$ and $\omega_p$. 
Such tuning is unlikely to occur in realistic astrophysical environments.  

Furthermore, we solved the general dispersion relation numerically without assuming either the MHD 
or non-MHD limits. These results confirm that no significant conversion of ALPs into photons occurs
under typical intergalactic conditions.  

 In conclusion, the perturbative analysis presented in this allows the possibility of conversion of ALPs into photons in the axion star. 
 Our analysis of the dispersion-relation shows that the condition under which a significant conversion ALPs into photons can take place
 if the condition $\frac{\beta^2}{m^2_a-\omega^2_p}\sim 1$ is satisfied.
 Since typical values of $\beta^2\sim$ and $\omega^2_p\sim $, a very high degree of fine tuning between $m^2_a$ and $\omega^2_p$ is required to satisfy the above condition. Thus, a star made of ALPs likely to remain stable in the magnetized intergalactic environment.
However, in extreme environments with very strong magnetic fields (such as
near magnetars \cite{manzari, PhysRevX.14.041015}), the situation may differ and deserves
further investigation.

\end{document}